\documentclass[longauth]{aa}  

\usepackage{graphicx}
\usepackage{txfonts}
\newcommand\fion[2]{$[$#1{\sc #2}$]$}

\begin{document} 

\title{The mysterious optical afterglow spectrum of GRB\,140506A at $z=0.889$
\thanks{Based on observations carried out under prog. ID 093.A-0069(B) with
the X-Shooter spectrograph installed at the Cassegrain focus of the Very Large
Telescope (VLT), Unit 2 -- Kueyen, operated by the European Southern
Observatory (ESO) on Cerro Paranal, Chile. Part of the observation were
obtained with Magellan as part of the program CN2014A-114.
}}


\author{
J.~P.~U. Fynbo\inst{1}
\and
T. Kr{\"u}hler\inst{2}
\and
K. Leighly\inst{3}
\and
C. Ledoux\inst{2}
\and
P.~M. Vreeswijk\inst{4}
\and
S. Schulze\inst{5,6}
\and
P. Noterdaeme\inst{7}
\and
D. Watson\inst{1}
\and
R. A. M. J. Wijers\inst{8}
\and
J. Bolmer\inst{9}
\and
Z. Cano\inst{10}
\and
L. Christensen\inst{1}
\and
S. Covino\inst{11}
\and
V. D'Elia\inst{12,13}
\and
H. Flores\inst{14}
\and
M. Friis\inst{10}
\and
P. Goldoni\inst{15}
\and
J. Greiner\inst{9}
\and
F. Hammer\inst{16}
\and
J. Hjorth\inst{1}
\and
P. Jakobsson\inst{10}
\and
J. Japelj\inst{17}
\and
L. Kaper\inst{8}
\and
S. Klose\inst{18}
\and
F. Knust\inst{9}
\and
G. Leloudas\inst{1}
\and
A. Levan\inst{19}
\and
D. Malesani\inst{1}
\and
B. Milvang-Jensen\inst{1}
\and
P. M\o ller\inst{20}
\and
A. Nicuesa Guelbenzu\inst{18}
\and
S. Oates\inst{21}
\and
E. Pian\inst{22}
\and
P. Schady\inst{9}
\and
M. Sparre\inst{1}
\and
G. Tagliaferri\inst{11}
\and
N. Tanvir\inst{23}
\and
C.~C. Th{\"o}ne\inst{21}
\and
A. de Ugarte Postigo\inst{21,1}
\and
S. Vergani\inst{16}
\and
K. Wiersema\inst{23}
\and
D. Xu\inst{1}
\and
T. Zafar\inst{20}
}
\institute{
Dark Cosmology Centre, Niels Bohr Institute, Copenhagen University, Juliane Maries Vej 30, 2100 Copenhagen O, Denmark\\
\email{jfynbo@dark-cosmology.dk}
\and
European Southern Observatory, Alonso de C{\'o}rdova 3107, Vitacura, Casilla 19001, Santiago 19, Chile
\and
Homer L. Dodge Department of Physics and Astronomy, The University of Oklahoma, 440 West Brooks Street, Norman, OK 73019, USA
\and
Department of Particle Physics and Astrophysics, Weizmann Institute of Science, Rehovot 76100, Israel
\and
Instituto de Astrof{\'i}sica, Facultad de F{\'i}sica, Pontificia Universidad Ca{\'o}lica de Chile, Av. Vicu{\~n}a Mackenna 4860, Santiago, Chile
\and
Millennium Institute of Astrophysics, Vicu\~{n}a Mackenna 4860, 7820436 Macul, Santiago, Chile
\and
Institut d'Astrophysique de Paris, CNRS-UPMC, UMR 7095, 98bis Bd Arago, 75014, Paris, France
\and
Astronomical Institute Anton Pannekoek, University of Amsterdam, PO Box 94249, NL-1090 GE Amsterdam, the Netherlands
\and
Max-Planck-Institut f{\"u}r extraterrestrische Physik, Giessenbachstrasse, 85748, Garching, Germany
\and
Centre for Astrophysics and Cosmology, Science Institute, University of Iceland, Dunhagi 5, 107 Reykjav{\'i}k, Iceland
\and
INAF / Brera Astronomical Observatory, Via Bianchi 46, 23807, Merate (LC), Italy
\and
INAF - Osservatorio Astronomico di Roma, via Frascati 33, 00040, Monteporzio Catone, Italy
\and
ASI-Science Data Center, Via del Politecnico snc, I-00133 Rome, Italy
\and
Laboratoire Galaxies Etoiles Physique et Instrumentation, Observatoire de Paris, 5 place Jules Janssen, 92195 Meudon, France
\and
APC, Astroparticule et Cosmologie, Univ. Paris Diderot, CNRS/IN2P3, CEA/Irfu, Obs. de Paris, Sorbonne Paris Cit{\'e}, 10 rue Alice Domon et L{\'e}onie Duquet, 75205, Paris Cedex 13, France
\and
Laboratoire GEPI, Observatoire de Paris, CNRS-UMR8111, Univ. Paris Diderot, 5 place Jules Janssen, F-92195 Meudon, France
\and
Faculty of Mathematics and Physics, University of Ljubljana, Jadranska ulica 19, SI-1000 Ljubljana, Slovenia
\and
Th{\"u}ringer Landessternwarte Tautenburg, Sternwarte 5, D-07778 Tautenburg, Germany
\and
Department of Physics, University of Warwick, Coventry CV4 7AL, UK
\and
European Southern Observatory, Karl-Schwarzschildstrasse 2, D-85748 Garching, Germany
\and
Instituto de Astrof{\'i}sica de Andaluc{\'i}a, Glorieta de la Astronom{\'i}a s/n, E-18008 Granada, Spain
\and
Scuola Normale Superiore di Pisa, Piazza dei Cavalieri 7, I-56126 Pisa, Italy
\and
Department of Physics and Astronomy, University of Leicester, University Road, Leicester LE1 7RH, UK
}
   \date{Received 2014; accepted, 2014}

 
  \abstract
{Gamma-ray burst (GRBs) afterglows probe sightlines to star-forming regions in distant star-forming galaxies. Here we 
present a study of the peculiar afterglow spectrum of the $z=0.889$ {\it Swift} GRB\,140506A.}
   {Our aim is to understand the origin of the very unusual properties of the
absorption along the line-of-sight.}
{We analyse 
spectroscopic observations obtained with the X-shooter spectrograph mounted on
the ESO/VLT at two epochs 8.8 h and 33 h after the burst as well as imaging
from the GROND instrument. We also present imaging
and spectroscopy of the host galaxy obtained with the Magellan telescope.}
   {The underlying afterglow appears to be a typical afterglow of a
long-duration GRB. However, the material along the line-of-sight has imprinted
very unusual features on the spectrum. Firstly, there is a very broad and
strong flux drop below 8000~\AA \ ($\sim$4000~\AA \ in the rest frame), 
which seems to be variable between the two
spectroscopic epochs. We can reproduce the flux-drops both as a giant 2175~\AA \ extinction bump
and as an effect of multiple scattering on dust grains in a dense environment.
Secondly, we detect absorption lines from excited \ion{H}{i} and \ion{He}{i}.
We also detect molecular absorption from CH$^+$.}
{We interpret the unusual properties of these spectra as reflecting the 
presence of three distinct regions along the line-of-sight: the excited
\ion{He}{i} absorption originates from an \ion{H}{ii}-region, whereas the Balmer 
absorption must originate from an associated photodissociation region. The 
strong metal line and molecular absorption and the dust extinction must originate from a
third, cooler region along the line-of-sight. The presence of (at least) three
separate regions is reflected in the fact that the different absorption 
components have 
different velocities relative to the systemic redshift of the host galaxy.}


   \maketitle
%

\section{Introduction}

Gamma-ray bursts (GRBs) have become a powerful tool to probe the interstellar
medium (ISM) of star-forming galaxies \citep[e.g.,][]{Jakobsson04,
2007ApJ...666..267P,Fynbo09,Kruhler13}. Optical spectroscopy of the afterglows
allows the measurement of a wide range of important properties of galaxies
such as chemical abundances \citep{2006NJPh....8..195S,2006A&A...451L..47F,
cct13,Sparre14},
molecular content \citep{2006A&A...451L..47F,2009ApJ...691L..27P, Kruhler13, 
DElia14, Mette14}, and dust
extinction \citep{Watson06,Li08,Ardis09,2009ApJ...691L..27P,Perley10,2011A&A...532A.143Z,Schady12,DeCia,Covino13}.

In this paper we present spectroscopic observations of the unusual 
afterglow of
the $z=0.889$ GRB\,140506A. GRB\,140506A was detected by {\it Swift} on 2014 
May 6
21:07:36 UT \citep{2014GCN..16214...1G}. The prompt emission was also detected
by the $Konus$ and $Fermi$ satellites \citep{2014GCN..16220...1J,
2014GCN..16223...1G}.  The burst was relatively long with a $T_{90}$ duration
of $111.1\pm9.6\rm$ s and the temporal profile is characterized by a sharp initial peak
with an extended tail \citep{2014GCN..16218...1M}. These properties
are fully within the range of normal long-duration GRBs, but the GRB optical 
afterglow has very unusual properties as we will discuss here.

The paper is organised in the following way: in Sect.~\ref{obs} we present 
our observations, in Sect.~\ref{res} our results and in Sect.~\ref{disc}
and Sect.~\ref{conclude}
we offer our discussion of the results and conclusions.

\section{Observations}
\label{obs}

On May 7 2014, with a mid exposure time of 8.8 h post burst, we acquired a
medium-resolution spectrum with the X-shooter spectrograph \citep{Vernet11}
mounted at the
ESO/VLT, covering a range 300 to 2300 nm \citep{2014GCN..16217...1F}.
The
spectrum was taken under excellent conditions with a photometric sky and a very
good seeing of 0\farcs54 in the $R$ band. The observation was carried out in
two executions of a 4$\times$600 s observing block following an ABBA nodding
pattern. The slit was re-aligned with the parallactic angle between the two
executions of the observing block.  The slit widths were 1\farcs0, 0\farcs9,
and 0\farcs9 in the UVB, VIS, and NIR arms, respectively. 
The airmasses at the start and end of the
spectroscopic observation were 1.43 and 1.22, respectively. For the given
instrument setup, the nominal resolution is 5500, 8800, and 5100 in the UVB,
VIS and NIR, respectively. As the seeing was smaller than the slit width the
actual resolution is higher than this. For the VIS, we measure the resolution from
the width of telluric absorption lines and find it to be 24 km
s$^{-1}$ (FWHM). For the UVB and NIR arms we find $\approx$ 40
and 42 km s$^{-1}$. All wavelengths are in vacuum and corrected for the 
heliocentric velocity of 18.9 km s$^{-1}$.

Because of very unusual features in the spectrum, we decided to observe the 
afterglow again the following night, 33 h after the burst, executing again two
4$\times$600 s observing blocks. The setup was the same as for the first
epoch spectrum and again with the slit aligned with the parallactic angle. 
The observing conditions were again excellent
with photometric conditions, a seeing of 0\farcs59 in the $R$-band, and an
airmass ranging from 1.27 to 1.21. 
The X-shooter spectra were reduced using the official ESO pipeline
version 2.2.0 \citep{Goldoni06,Modigliani10}

The Gamma-Ray burst Optical Near-Infrared Detector (GROND,
\citealp{2008PASP..120..405G}) mounted on the MPG 2.2m telescope on La Silla
observed the field of GRB~140506A on 2014-05-09,
2014-05-10, and 2014-05-28. A late-time host galaxy observation was
carried out on 2014-07-14. 
Simultaneous photometry was obtained in seven
broad-band filters ($g'r'i'z'$ in the optical and $JHK_{\rm s}$ in the
near-infrared wavelength range) with a total integration time of around 90 mins
in $g'r'i'z'$ and 75 mins in $JHK_{\rm s}$ on each visit. We reduced and
analyzed the data using custom-made software largely based on pyraf/IRAF
\citep{Tody1993}, closely following the procedure outlined in
\citet{Tom08}. Photometric calibration was performed against
field stars from the 2MASS catalog \citep{2006AJ....131.1163S} in the NIR. The
photometry of the optical bands was tied to observations of an SDSS
\citep{2011ApJS..193...29A} standard field taken immediately before the GRB
field. We also use these calibration data for the X-shooter acquisition camera
and the Magellan/IMACS observations. Details of our photometry are given in
Table~\ref{tab:Phot}.

\begin{table}
\caption{Photometry of the afterglow and host galaxy from May, 
June and July 2014\label{tab:Phot}}
\centering
\begin{tabular}{cccc}
\hline
\hline\noalign{\smallskip}
{Instrument} & {Filter} & UT Start time & Magnitude\tablefootmark{a,b} \\
             &          &               &  (mag) \\

\hline\noalign{\smallskip}
X-shooter & $R$      & 2014-05-07 05:03 &  $21.40\pm0.08$ \\
X-shooter & $R$      & 2014-05-07 05:19 &  $21.44\pm0.11$ \\
X-shooter & $R$      & 2014-05-07 06:00 &  $21.69\pm0.11$ \\
X-shooter & $R$      & 2014-05-08 06:00 &  $22.39\pm0.09$ \\

GROND & $g'$         & 2014-05-09 07:18 &  $23.73\pm0.05$ \\
GROND & $r'$         & 2014-05-09 07:18 &  $22.57\pm0.04$ \\
GROND & $i'$         & 2014-05-09 07:18 &  $22.11\pm0.05$ \\
GROND & $z'$         & 2014-05-09 07:18 &  $21.82\pm0.06$ \\
GROND & $J$          & 2014-05-09 07:18 &  $21.09\pm0.13$ \\
GROND & $H$          & 2014-05-09 07:18 &  $20.88\pm0.18$ \\
GROND & $K_{\rm s}$  & 2014-05-09 07:18 &  $20.82\pm0.22$ \\

GROND & $g'$         & 2014-05-10 09:17 &  $23.86\pm0.19$ \\
GROND & $r'$         & 2014-05-10 09:17 &  $22.76\pm0.11$ \\
GROND & $i'$         & 2014-05-10 09:17 &  $22.37\pm0.11$ \\
GROND & $z'$         & 2014-05-10 09:17 &  $22.16\pm0.14$ \\
GROND & $J$          & 2014-05-10 09:17 &  $>20.95$ \\
GROND & $H$          & 2014-05-10 09:17 &  $>20.45$ \\
GROND & $K_{\rm s}$  & 2014-05-10 09:17 &  $>19.48$ \\

GROND & $g'$         & 2014-05-28 04:59 & $24.71\pm0.11$ \\
GROND & $r'$         & 2014-05-28 04:59 & $24.24\pm0.10$ \\
GROND & $i'$         & 2014-05-28 04:59 & $23.66\pm0.17$ \\
GROND & $z'$         & 2014-05-28 04:59 & $23.13\pm0.17$ \\
GROND & $J$          & 2014-05-28 04:59 & $>21.98$ \\
GROND & $H$          & 2014-05-28 04:59 & $>21.38$ \\
GROND & $K_{\rm s}$  & 2014-05-28 04:59 & $>20.43$ \\

IMACS & $white$      & 2014-06-26 03:04 & $24.27\pm0.14$ \\

GROND & $g'$         & 2014-07-14 03:53 & $>24.8$ \\
GROND & $r'$         & 2014-07-14 03:53 & $24.43\pm0.21$ \\
GROND & $i'$         & 2014-07-14 03:53 & $23.70\pm0.15$ \\
GROND & $z'$         & 2014-07-14 03:53 & $23.71\pm0.23$ \\
GROND & $J$          & 2014-07-14 03:53 & $>22.07$ \\
GROND & $H$          & 2014-07-14 03:53 & $>21.45$ \\
GROND & $K_{\rm s}$  & 2014-07-14 03:53 & $>20.53$ \\
\hline\noalign{\smallskip}

\end{tabular}
\tablefoot{\tablefoottext{a}{All photometry is given in the AB system, and is
not corrected for Galactic foreground reddening.}\tablefoottext{b}{Upper
limits at 3$\sigma$ confidence level.}}
\end{table}

Magellan observations were done in the following way.
We first obtained a deep image of the field with the f/4 camera on the
Inamori-Magellan Areal Camera and Spectrograph (IMACS; \citealt{Dressler2011a})
on the Magellan/Baade 6.5-m telescope. The observation started at 03:04 UT
on June 26, 2014 (i.e. 51.5 days after the GRB) and comprised of three
unfiltered images with an individual exposure time of 300 s.

\begin{figure}
\centering
\includegraphics[width=8.0cm,clip]{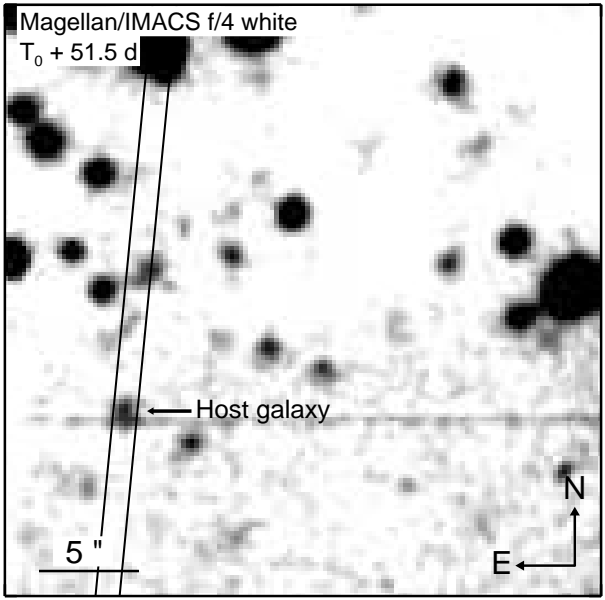}
\caption{
Field of GRB 140506A ($30''\times30''$). The position of the host galaxy is
marked.  The slit of the late-time spectrum was inclined by $-5\fdg5$ East
of North. The
image has been smoothed with a Gaussian kernel to increase the visibility 
of the host
galaxy.
}
\label{host}
\end{figure}

After the host galaxy was detected in the combined image, we acquired a spectrum
starting 03:47 UT, comprising of six exposures with an individual integration
time of 1200~s (see Fig.~\ref{host}). At the redshift of the GRB, the emission
lines lie in the red part of the visual spectrum. To disentangle them from the
sky emission lines, we chose an intermediate-resolution grating with 600
lines/mm and a blaze angle of 15\fdg4.  For the given configuration, the
dispersion is 0.39~\AA/px, the resolving power is about 3600 and the spectral range
extends from 6600~\AA \ to 9770~\AA. We chose a slit width of 1\farcs2 to match
the seeing conditions.  The Magellan data were reduced and calibrated using
standard procedures in IRAF \citep{Tody1993}. The extraction aperture had a
width of 10 pixels (2\farcs2). The spectrum was flux-calibrated with the
spectrophotometric standard star Feige 56. The wavelength solution was
transformed to vacuum wavelengths.

\section{Results}
\label{res}

\subsection{Absorption and emission lines}

\begin{figure}
\centering
\includegraphics[width=8.2cm,clip]{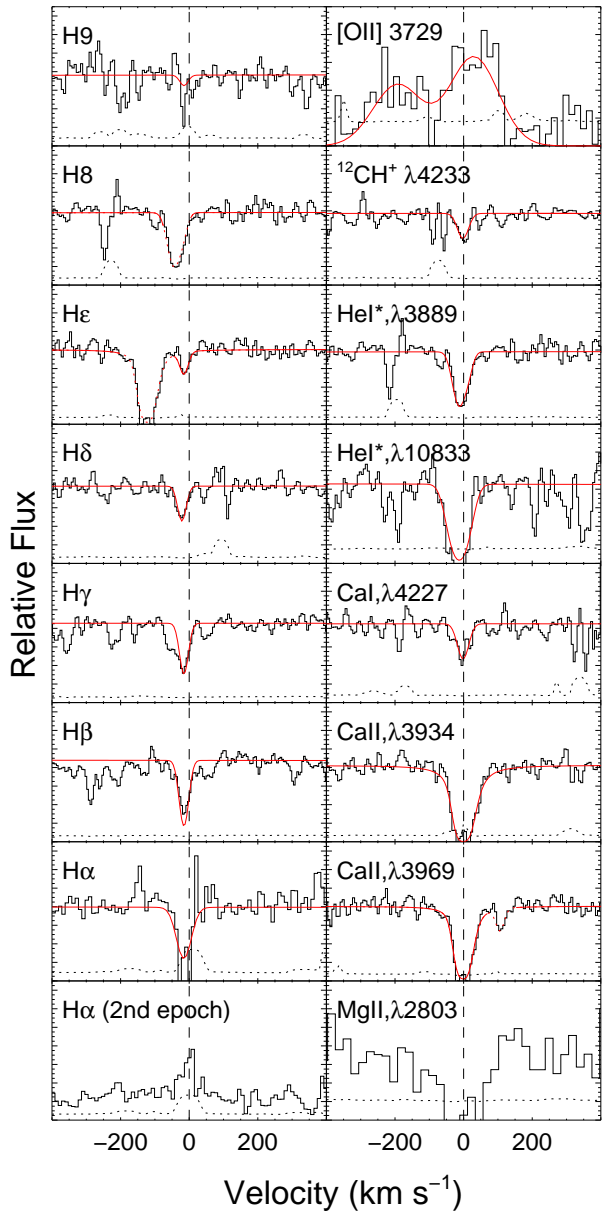}
\caption{
The main absorption and emission lines in the spectrum.
The noise spectra are
overplotted with a dotted line and the plot range on the y-axis starts at zero.
The scale on the y-axis is arbitrary. The zero point for the velocity scale
is set to $z=0.88911$.
 In the left column we show the Balmer
lines that are centred at $-14$ km s$^{-1}$ ($z=0.88902$). 
For H$\alpha$ we show the region around 
the line from both X-shooter spectra. In the right column we show \ion{Ca}{ii}
and \ion{Ca}{i},
the \ion{He}{i}* lines centred at $-11$ km s$^{-1}$ ($z=0.88904$) and the CH$^+$ line
centred at the same velocity as the \ion{Ca}{ii} lines.  We also show the
\fion{O}{ii} doublet, which is redshifted by 30 km s$^{-1}$ relative to
\ion{Ca}{ii} and have a velocity width of 150 km s$^{-1}$. Note that H8 is
blended with \ion{He}{i}*$\lambda3889$. Also overplotted are Voigt profile
fits to the absorption lines (except \ion{Mg}{ii} that is in the UVB
part of the spectrum where the signal-to-noise ratio is too low) and a 
Gaussian fit to the \fion{O}{ii} doublet. For the \ion{Ca}{ii} lines we 
show the Voigt-fits with only a single absorption component.
}
\label{abslines}
\end{figure}

\begin{figure}
\centering
\includegraphics[width=8.2cm,clip]{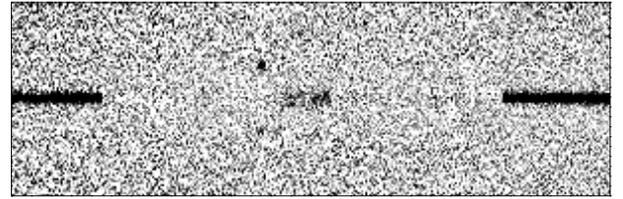}
\caption{
The \fion{O}{ii} emission line doublet from the
underlying host galaxy. The afterglow has been subtracted assuming that the
afterglow spectrum can be approximated by a linear function across the position
of the line (the compact dot above the trace is a cosmic ray hit).  Based
on the line we infer a star-formation rate of 1 $M_{\odot}$ yr$^{-1}$.
}
\label{hostOII}
\end{figure}

Based on absorption lines from \ion{Fe}{ii}, \ion{Mg}{ii}, \ion{Ca}{ii},
\ion{He}{i}* and \ion{H}{i}* we infer a redshift of 0.889 for the burst
(Fig.~\ref{abslines}). This is confirmed by the detection of \fion{O}{ii}
emission from the underlying host galaxy (Fig.~\ref{abslines}, top right panel,
and Fig.~\ref{hostOII}.) redshifted by 30 km s$^{-1}$ relative to the
\ion{Ca}{ii} absorption lines. From the strength of the \fion{O}{ii} we infer a
star-formation rate (SFR) of 1 M$_{\sun}$ yr$^{-1}$ using the relation between
\fion{O}{ii} luminosity and SFR in \citet{Kennicutt}. An SED analysis of 
the host galaxy based on the host photometry in Table~\ref{tab:Phot} following
the procedure in \citet{Tom11} yields that the host appears to be a typical 
GRB host
with a stellar mass of $\sim$10$^9 M_{\sun}$, A$_{V} \sim 1$
and with a dust-corrected SFR of a few solar masses per year. As seen in
Fig.~\ref{host} the host is also detected in the Magellan white light image.

The Magellan spectrum is unfortunately hampered by strong pick-up noise.
We detect the \fion{O}{ii} emission line at low signal-to-noise ratio 
($\lesssim3\sigma$), but no other 
lines are securely detected. 

\begin{table}
\caption{Observer frame Equivalent widths for detected lines
at both epochs.}
\label{table:ew}
\centering
\begin{tabular}{l c c c l l l }     
\hline\hline
Line &  EW$_1$ & EW$_2$ \\
   &  (\AA)    & (\AA)   \\
\hline
\ion{Fe}{ii}$\lambda2600$ & 2.2$\pm$0.9    & 2.9$\pm$1.1 \\
\ion{Mg}{ii}$\lambda2796$ & 3.2$\pm$0.5    & 1.9$\pm$0.7 \\
\ion{Mg}{ii}$\lambda2803$ & 3.7$\pm$0.5    & 2.0$\pm$0.7 \\
\ion{He}{i}*$\lambda3889$ & 1.13$\pm$0.11  & 1.4$\pm$0.3 \\
\ion{Ca}{ii}$\lambda3934$ & 2.10$\pm$0.11  & 2.5$\pm$0.3 \\
\ion{Ca}{ii}$\lambda3969$ & 1.91$\pm$0.11  & 1.6$\pm$0.3 \\
H$\epsilon$                & 0.22$\pm$0.05  &   --        \\
H$\delta$                  & 0.40$\pm$0.05  & 0.0$\pm$0.2 \\
H$\gamma$                  & 0.86$\pm$0.05  & 0.3$\pm$0.2 \\
H$\beta$                   & 0.73$\pm$0.05  & 0.3$\pm$0.2 \\
H$\alpha$                  & 1.6$\pm$0.3  & $-$4.7$\pm$1.2 \\
\ion{He}{i}*$\lambda10833$ & 5.2$\pm$0.5   & 4.9$\pm$1.5 \\
$^{12}$CH$^+$$\lambda4233$ & 0.37$\pm$0.05  & 0.38$\pm$0.10 \\
\fion{O}{ii}               & $-$1.59$\pm$0.13 & $-$3.3$\pm$0.3 \\
\hline
\end{tabular}
\end{table}

\ion{He}{i}* and Balmer line absorption has to our knowledge never been detected
before in a GRB optical afterglow spectrum.  The measured equivalent widths
(EWs) of the main absorption and emission lines are provided in
Table~\ref{table:ew}. The ratio between the EWs of the
\ion{He}{i}*$\lambda$10833 and \ion{He}{i}*$\lambda$3889 lines is 4.6$\pm$0.5.
The expected ratio for unsaturated lines from atomic physics is 23.3 indicating
significant saturation of at least the \ion{He}{i}*$\lambda$10833 line.  There
is evidence for variability of the strength of the Balmer absorption lines.  We
suspect that this is due to blending with the underlying host emission lines for the
following reasons. Firstly, in Table~\ref{table:ew} we give the EWs of the
\fion{O}{ii} doublet showing that the change in the EWs of emission lines
easily are strong enough to cause such an effect. Secondly, the estimated
strength of the Balmer emission lines given the SFR would be strong enough.
Finally, in the second epoch spectrum there is evidence for H$\alpha$ emission
(see the lower left panel of Fig.~\ref{abslines}).  Unfortunately H$\alpha$ is
located on a sky line and our Magellan spectrum is not deep enough so we need a
deeper spectrum of the host to be certain about this interpretation of the
Balmer line variability.

Finally, we also detect an absorption line consistent with the 4233~\AA \
line from CH$^+$. There is also a weaker feature consistent with the 3958~\AA \
line. We do not detect absorption from CH. In the MW
CH$^+$ is sometimes stronger than CH \citep[e.g.,][]{Smoker14}
so the detection of CH$^+$ and nondetection of CH is not unexpected.

To derive column densities we have carried out Voigt-profile fits of
the Balmer, \ion{He}{i}*, \ion{Ca}{ii}, \ion{Ca}{i} and CH$^+$ lines.
The results of these fits are provided in Table~\ref{table:voigt}. 
To obtain acceptable fits we have to adopt different velocity centroids for the
Calcium and CH$^+$ lines ($z=0.88911$), the Balmer lines ($z=0.88902$) and the
excited Helium ($z=0.88904$). The uncertainties on the redshifts are 0.00002 so
the latter two redshifts are marginally consistent. We then tie the Doppler
parameters of lines at the same redshifts together and obtain
$b=16.5\pm8.0\,\mathrm{km\,s}^{-1}$ for Calcium and CH$^+$,
$b=8.5\pm1.5\,\mathrm{km\,s}^{-1}$ for the Balmer, and
$b=18.5\pm3.4\,\mathrm{km\,s}^{-1}$ for the Helium lines respectively. These
fits assume a single component per ion/molecule.

For the very strong \ion{Ca}{ii}($\lambda\lambda 3934,3969$)
transitions, in principle, many components with a low column density that are
not seen in the weaker lines could contribute to the observed line shape. In an
attempt to resolve this ambiguity between total column density and number of
components, we fitted the \ion{Ca}{ii} in such a way that the total number of
components was a free parameter in the fit until adding more components did not
provide an improvement in the fit statistics. For five unresolved components,
for example, the total \ion{Ca}{ii} column density would be $\log (N/
\mathrm{cm}^{-2}) = 16.5 \pm 0.3$. Fits with a fixed number of components
(between one and ten) return column densities between $\log (N /
\mathrm{cm}^{-2}) \sim 15.8$ and $\log (N / \mathrm{cm}^{-2}) \sim 17.0$.  Given
the medium resolution of X-shooter and moderate signal-to-noise of our spectra,
we can not make stronger statements on the column density of \ion{Ca}{ii}.

Voigt profile fits to the second epoch spectrum return within errors
consistent results except for the Balmer lines (see also Table~\ref{table:ew}).
Due to the lower signal-to-noise of the second epoch spectrum, however, these
fits are not further constraining. The decreasing equivalent width of the
Balmer lines in absorption is likely caused by an increasing contribution of
the Balmer lines in emission from the host galaxy.

We have searched for Diffuse Interstellar Bands \citep[e.g.,][]{Xiang11},
but we do not detect any of these above detection limits of $\gtrsim300$ m\AA
\ (3$\sigma$).

\begin{table}
\caption{Results from Voigt-profile fitting to the lines in the first epoch spectrum.}
\label{table:voigt}
\centering
\begin{tabular}{l c c c c c c }     
\hline\hline
Element & $z$ & $v$ & $b$ & $\log N$ \\
        &  & (km s$^{-1}$)  & (km s$^{-1}$) & ($\log \mathrm{cm}^{-2}$)  \\
\hline
\ion{H}{i}*  & 0.88902 & $-14$ & 8.5$\pm$1.5 & 13.96$\pm$0.12 \\
\ion{He}{i}* & 0.88904 & $-11$ & 18.5$\pm$3.4 & 13.99$\pm$0.12 \\
\ion{Ca}{ii} & 0.88911 & 0 & -- & $>16.5$ \\
\ion{Ca}{i} & 0.88911 & 0 & 16.5$\pm$8.0  & 12.07$\pm$0.17 \\
CH$^+$      & 0.88911 & 0 & 16.5$\pm$8.0  & 14.45$\pm$0.18 \\
\hline
\end{tabular}
\end{table}

\subsection{The peculiar SED}

\begin{figure*}[ht]
\centering
\includegraphics[width=17cm,clip]{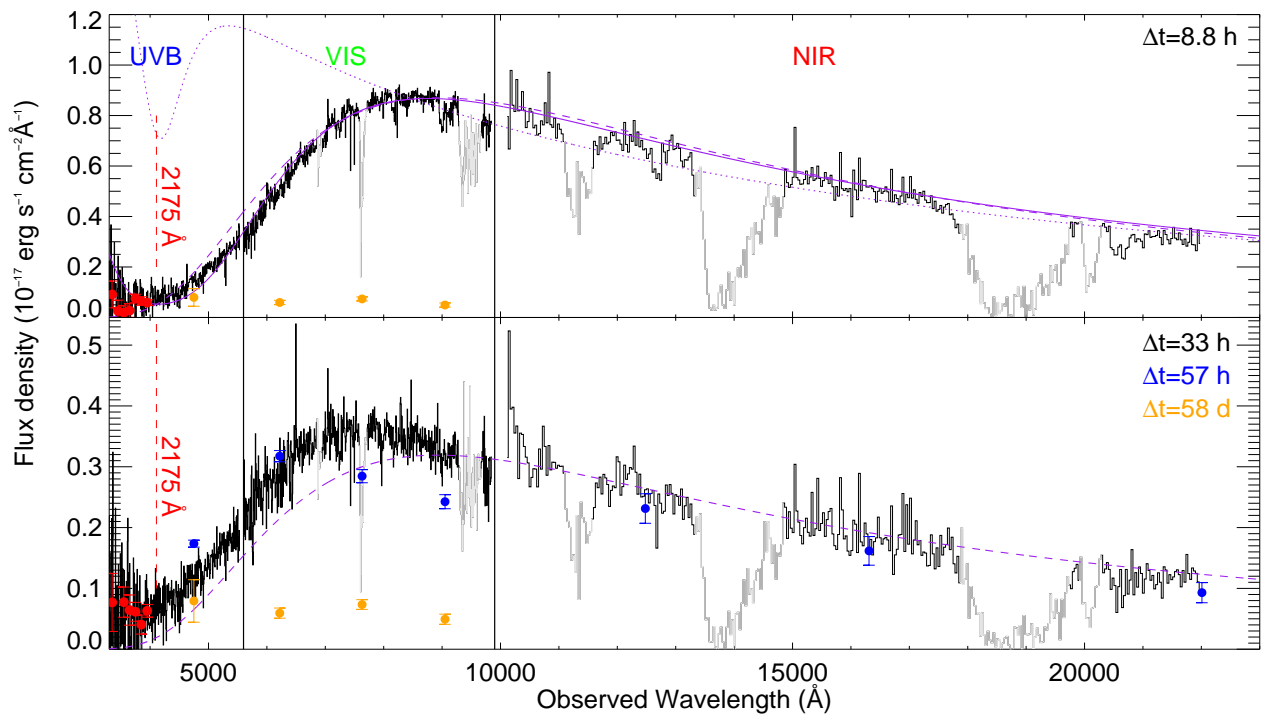}
\caption{
The X-shooter spectra taken 8.8 h (top) and 33 h (bottom) after the burst.  The
spectra are binned for better visibility.  The red points show the spectrum at
$\lambda < 4000$~\AA \ binned in 100~\AA \ bins.  Regions affected by telluric
absorption are plotted in light gray.  Also overplotted on the first epoch
spectrum is a power-law (with slope set by a fit to the X-ray afterglow and
assuming a $\Delta \beta = 0.5$ cooling break between the X-rays and the
optical) reddened by three different prescriptions: MW extinction curve (dotted
line), extreme 2175~\AA \ extinction bump (solid line), and extinction with
multiple scattering (dashed line).  The shape of the second epoch spectrum is
overall similar to the spectrum from the first epoch, but the flux drop occurs
at slightly bluer wavelengths. To make this more clear we rescale and overplot
the modeling of multiple scattering from the first epoch on the spectrum of the
second epoch. Furthermore, we show with blue points the GROND photometry from
about 57 h after the burst corrected for host galaxy emission and with
orange points the host photometry from 58 days after the burst.
}
\label{afterglowspec}
\end{figure*}

The spectral energy distribution (SED) is very unusual
(Figure~\ref{afterglowspec}): in the NIR and red part of the VIS arm the X-shooter
spectrum looks like a typical GRB afterglow, i.e. a power-law spectrum. Then it
breaks around 8000~\AA \ (rest frame 4000~\AA) and vanishes in the UVB. In
Figure~\ref{afterglowspec} we show the UVB spectrum at $\lambda_\mathrm{obs} 
< 4000$~\AA \
binned more heavily with red points with error bars. These points indicate a
rising trend in the bluest end of the spectra, especially in the second epoch.
The flux level in this bluest region of the spectrum is so low ($\sim$25 mag on
the AB system) that the host galaxy flux is likely to contribute a large
fraction of the detected emission. This is clear from 
Figure~\ref{afterglowspec} where it can be seen that the flux density of the
host galaxy continuum (as traced by the late time GROND photometry shown in
orange) is very 
similar to the flux density in the bluest part of the X-shooter spectra 
(as shown with red points).
However, without more information about the strength
of the host emission below 4000~\AA \ it is not possible to judge with 
certainty whether
afterglow emission contributes significantly to the X-shooter spectra below
$\sim4000$~\AA \ in the observed frame.

We also reanalysed the imaging data from UVOT.
The UVOT photometry shows detections of the afterglow securely in white light
and in the $U$, $B$ and $V$ filters \citep[see also][]{Siegel} during the 
first few hours after the burst. There is also a 
tentative detection (2.8$\sigma$) in the $UVW1$ band, but only during the first 
10 minutes after the burst. Overall, the UVOT photometry is consistent with the
shape of the SED inferred from the first epoch X-shooter spectrum.

It is
important to stress that the X-shooter spectra were taken in excellent conditions so the
flux drop in the blue is intrinsic to the event and is not caused by slit-loss.
In the following we explore different possible interpretations of this
peculiar SED.

\subsubsection{A giant 2175-\AA \ extinction bump}

The location of the flux-drop is in the rest-frame ultraviolet and it is therefore
natural to explore whether this can be an extreme example of the 2175~\AA \
extinction bump known from the Milky Way and also previously seen in GRB afterglow
spectra \citep{Ellison06,Fynbo07,Tom08,Ardis09,2009ApJ...691L..27P,2011A&A...532A.143Z}.
This feature is ubiquitous in the MW and in M31. It is also present along
sightlines in the LMC and at least one sightline in the SMC \citep[see the 
extensive discussion of this extinction feature in][]{Ardis09}.

The flux drop can be well matched by a \citet{FM07} parametrization of the
extinction curve except in the bluest region below $\sim4000$~\AA. 
An example is plotted with a solid line in Fig.~\ref{afterglowspec} with these parameters: 
c$_1$ = 1.0,
c$_2$ = 0.8,
c$_3$ = 115.,
c$_4$ = 0.46,
c$_5$ = 5.9,
$\gamma$ = 2.75,
R$_{V}$ = 3.1,
x$_\mathrm{c}$ = 4.6,
A$_{V}$ = 0.9. 
We here assume that the intrinsic spectrum is a power-law $F_{\nu} \propto
\nu^{-\beta}$, with $\beta=0.75$ as derived from the XRT spectrum ($\beta=0.75\pm0.07$)
and with a $\Delta\beta=0.5$ break between the X-rays and optical bands.
For consistency we have verified that the afterglow lightcurve
in the optical and x-ray bands are consistent with a scenario where the 
cooling break is located between the optical and X-ray bands.
The extinction bump is characterized
by c$_ 3$ and $\gamma$. \citet{FM07} introduce the bump height $E_\mathrm{bump} =
c_3/\gamma^2$ and along Milky Way (MW) sightlines $E_\mathrm{bump}$ and $\gamma$ are
in the range 1--6 and 0.8--1.5, respectively. For GRB\,140506A the values are
15 and 2.75, i.e. both are much larger than what is seen along any sightline in the
MW.

For comparison we also overplot with a dotted line a \citet{Pei1992}
parametrization of the MW extinction curve. Here we have assumed the
same underlying power-law spectrum and A$_{V}$ = 0.8. As seen it is here
impossible to match the depth and width of the bump. This illustrates
well how different the 2175~\AA \ bump has to be from that of the 
MW.

\subsubsection{Extinction with multiple scattering}

The extinction feature is also similar to what is seen in rare cases of 
reddened supernovae and active galactic nuclei \citep{Fynbo13,Karen14,Amanullah14}. 
These systems have been successfully modeled with the prescription
for multiple scattering of light discussed in \citet{Wang05} and 
\citet{Goobar08}.
The point is here that if the mean free path for dust scattering is smaller 
than the spatial extent of the dust cloud, the photon propagation is 
diffusive and the normal single scattering approximation is invalid. 
The dashed line in Fig.~\ref{afterglowspec} shows a simple model based on this
diffusive photon propagation described by the parameters
 $_\mathrm{V} = 0.65$, $p = -2.5$ and $a=0.8$, which are the relevant 
parameters for multiple scattering in the case of an underlying LMC-like
extinction curve \citep{Goobar08}. It is not possible to get a good fit
with the parameters relevant for an SMC-like extinction curve 
($a=1.0$, $p  = -1.07$; Goobar, private communication). 

Strictly, the approximation described in \citet{Goobar08} assumes an
isotropically emitting source.  As the emission from the GRB afterglow is
jetted the details of the effect most likely will be different. It is beyond
the scope of the present work to develop a more detailed model of the effective
extinction curve in the case of multiple scattering for a jetted source.

\subsubsection{SED variation}

The shape of the flux drop in the blue end of the spectrum changes between the
two epochs. In the lower panel of Fig.~\ref{afterglowspec} we overplot the
modeling of multiple scattering from the first epoch on the spectrum of the
second epoch which clear shows the change. The shape of the second epoch spectrum
is overall similar to the spectrum from the first epoch, but the flux drop
occurs at slightly bluer wavelengths. We also 
overplot on the second epoch spectrum 
the GROND photometry from about 57 h after the burst indicating an even weaker 
flux drop at this time.

A contributing cause to this is SED variation could be a changing relative
contribution from the underlying host galaxy, but as the host galaxy only
contributes 5 and 16\% of the total light in the $r$-band (see
Table~\ref{tab:Phot}) the host galaxy emission is insufficient to explain the
full variation. We return to this point below.

\section{Discussion}
\label{disc}

\subsection{Origin and variability of the flux drop}

The most remarkable property of the afterglow spectrum is the strong,
apparently variable flux drop bluewards of about 8000~\AA \ in the observer
frame. Whereas some of the variation of the shape of the flux drop may be
explained by an increasing host contribution, in particular from the second
epoch spectrum to the GROND photometry the following night, the host is too
faint at 24.4 mag in the $r$ band to explain the variation between our two
spectroscopic epochs (see afterglow magnitudes in Table~\ref{tab:Phot}).  More
specifically, we concluded this by establishing that the second epoch spectrum
cannot be reproduced by a linear combination of the first epoch spectrum and a
host galaxy model matching the GROND host photometry. 

Whatever causes the blue drop, it has a high optical depth; the most likely
cause of variation of the blue feature from 8 h to 33 h is then variation of
that optical depth perpendicular to the line of sight, combined with a 
significant growth of the emitting area we see, as explored for GRB\,021004
by \citet{Starling05}. To illustrate this, we use the standard expressions
for the expansion of the GRB blast wave in the pre-jet break phase
\citep[e.g.][]{WG99,Bing04}. Since material close to the GRB will be
severely affected by its radiation, we assume the absorber is many parsecs
from the explosion and does not interact with the blast wave.
This is consistent with distances inferred from variable fine-structure
lines for other bursts \citep[e.g.,][]{Vreeswijk07,DElia07,Vreeswijk11,DeCia12}.
The X-ray and UV afterglow may destroy dust out to $\sim$100 pc,
but this is far from always the case as discussed in \citet{Fruchter01}.
But due to the
blast wave slowing down, relativistic beaming decreases strongly 
and we see the emitting area expand: its size scales as $(\gamma ct)^2$,
and since in  this phase $\gamma\propto t^{-3/8}$, as $t^{5/4}$.
Between 8h and 33h it therefore expands by about a factor 6. The fraction
of flux that penetrates the absorber near the centre of the feature is
at most 5--10\% at both epochs.
This means that the absorber is big enough still to cover at least 90\% of the
emitting region at the second epoch, or it is even bigger and the
transmitted fraction merely indicates that the optical depth is a few,
so a small percentage of the light filters through. The blueward shift of
the turnover could either indicate a decrease in average optical depth of
the absorber while the area illuminated by the photosphere grows or, less 
likely, some light reaching us unabsorbed via a small fraction of ``holes''
in the absorber.

The alternative explanation of an extreme 2175-\AA \ bump appears less likely
as a similar strong bump is not observed anywhere in the MW \citep{FM07}. 

We note that \citet{Fynbo09} report an unexplained flux drop in the blue of
the afterglow of GRB\,070318. This event is discussed in more detail in 
\citet{Watson09}.
This flux drop can also be well reproduced 
by the \citet{Goobar08} multiple scattering scenario. 

We note that steep UV extinction, albeit less extreme than what we see here, is
also seen towards the MW bulge, the central region of M31 and towards several
AGN \citep[e.g.,][]{Fynbo13,Nataf13,Jiang13,Dong14,Karen14}.

\subsection{Absorption lines}

The metal lines detected in the spectrum are strong, but not unusually strong
for a GRB afterglow. The line-strength parameter defined in \citet{Antonio12}
is $-$0.11$\pm$0.15, which implies average line-strengths compared to the 
sample of \citet{Fynbo09}. 

In contrast to this, the absorption lines from excited hydrogen and helium 
are very unusual. To our knowledge, they have never been detected before in a GRB 
afterglow spectrum and 
they are not detected in the composite afterglow
spectrum presented in \citet{Lise} down to a limit of about 80 m\AA\ for H$\beta$
and \ion{He}{i}$^*$$\lambda$3889 (3$\sigma$). 
We have explored whether the population of the excited states of H and He can
be explained by the GRB afterglow UV photons exciting a neutral absorption
cloud at a distance of 50-500~pc away from the burst. Such an UV pumping model
can naturally explain the observations of excited states of ions such as \ion{Fe}{ii}
and \ion{Ni}{ii} observed along several GRB sightlines \citep{Prochaska06,Vreeswijk07,DElia07}.
We closely follow the methodology of \citet{Vreeswijk13}, which includes both
excitation and ionization of the relevant ions (H and He in this case) and a
solar helium abundance is assumed.  Varying the distance and the initial
neutral hydrogen column density over a reasonable range, we find that the
population of the excited state of H does not reach above $10^{11}$ atoms
cm$^{-2}$, while the observed value is almost $10^{14}$ cm$^{-2}$.  For HeI$^*$
the difference is less dramatic, but the modeled population is still at least a
factor of 10 below the observed value. Therefore, we conclude that excitation
by GRB afterglow photons of a nearby neutral absorber is not a viable
explanation for the presence of the Balmer and HeI$^*$ absorption
lines.  

It might be that a dense absorber is situated close to the GRB, and is being
ionized by it \citep[see also][]{LP02,Prochaska08,KP13}.  In many 
astrophysical environments, including quasars and AGN, \ion{He}{1}* is
populated by recombination, rather than pumping
\citep[see, e.g.,][for a quasar application]{leighly11}.  In that case,
we would expect to see \ion{He}{i}* only if sufficient He$^+$ could
recombine to the $n=2$ level after the GRB.  The timescale for
recombination to the triplet state for helium is $\sim 1.5\times10^5/n_e\rm \,
years$, and hence even for very high densities $n$ it is difficult to
understand how the observed column densities can be achieved before
the time of the first X-shooter spectrum.  

A third explanation is that the sightline happens to
intersect a region in the host galaxy with the observed column densities,
caused by an ionizing source different from the GRB, e.g. a massive star or
cluster of stars \citep[see also][]{Darach13,KP13}. 
Absorption lines from excited \ion{He}{i}* are
detected in the ISM of young clusters in the Milky Way. The lines are here
formed in \ion{H}{ii} region around the hot stars.  The EW of the
\ion{He}{i}*$\lambda3889$ line measured in the interstellar clouds in front of
the Trapezium in Orion is 0.12~\AA \ \citep{Oudmaijer1997}. The strength of the
\ion{He}{i}*$\lambda3889$ line towards GRB\,140506A is hence 5 times stronger
than that seen towards the Trapezium in Orion.  The EW of the
\ion{He}{i}*$\lambda10833$ line is not given in \citet{Oudmaijer1997}, but
judging from their Fig.~6 it is  3-5 times higher than for the
\ion{He}{i}*$\lambda3889$ line, i.e. a similar ratio to what we see.
\citet{Evans2005} report that \ion{He}{i}*$\lambda3889$ is seen towards many
stars in their study of three clusters (NGC 6611, NGC 3293 and NGC 4755)
that still contain O and/or B stars,
but with large spatial variations.
In addition, the strong absorption often detected in the X-ray 
afterglows of GRBs has been argued to originate from ionized helium in the 
natal \ion{H}{ii} of the GRB progenitor star \citep{Darach13}.

To explore this scenario we have performed photoionization modeling using Cloudy \citep{ferland13}
roughly appropriate for a large \ion{H}{ii} region and accompanying 
photo-dissociation region (PDR) illuminated by a star cluster.  We
used a spectral energy distribution consisting of a 5 Myr starburst
\citep{bc03} with added soft X-ray emission to approximate the the
emission of the O stars \citep{sciortino90}.  We assumed constant
pressure, and used solar abundances with no dust.  We found that we
were able to easily reproduce the observed \ion{He}{i}* column
density in the \ion{H}{ii} region of illuminated gas slab, as
expected, provided that the ionization parameter was high enough
\citep[e.g.,][]{leighly11}.  The Balmer absorption occurred in the
partially ionized zone/photodisscociation region beyond the Str{\"o}mgren
sphere.  The Balmer absorption could only be matched if there is a
very high density at the illuminated face, around $\log{n / \mathrm{cm}^{-3}} =7.75$ 
corresponding to $ \log{n/ \mathrm{cm}^{-3}} \sim 8.25$ in the Balmer-line absorbing
gas; when the density is high, Lyman-$\alpha$ scattering populates
the first excited state in hydrogen. These values are much higher than 
typical for an
Galactic \ion{H}{ii} region ($n \sim 10^3\rm \, cm^{-3}$), but such
densities are found, e.g., in ultra compact \ion{H}{ii} regions
\citep{Churchwell02} and photodisscociation regions \citep{HT97}. 
The separation
between the location of the absorption for the different ions explains
the different line widths.  

The possible absorption line from CH$^+$ is strong, but given the correlation
with the hydrogen column density seen in the MW 
\citep[][their Fig. 3]{Smoker14} it is not unreasonably strong given the
very large hydrogen column densities typically probed by GRB sightlines
\citep{Jakobsson06,Fynbo09}.

\subsection{Implications for dark GRBs}

It is intriguing that if this afterglow had been located at a more typical 
redshift of $z\sim2$ the flux drop would have been located in the observed
$J$ band and the burst would have been extremely dark in the observed optical
band despite being located behind a relatively modest amount of dust with 
$A_{V} \sim 1$. Hence, the effect that causes the unusual spectral
shape in the afterglow of GRB\,140506A may be important for the understanding 
of the dark burst phenomenon \citep{Fynbo01,Palli04,Greiner11,Tom11,Rossi12,Perley13}.

\section{Conclusions}
\label{conclude}

In summary we have analysed the afterglow spectrum of GRB\,140506A.
The spectrum has two peculiar properties: {\it i)} an unusual extinction signature
with the very strong and variable flux-drop below 8000~\AA \ (4000 \AA \
in the restframe), and {\it ii)} absorption lines from exited hydrogen and
helium. 
In our view the most likely scenario consistent with the observations is a 
sightline passing through several distinct regions. The absorption from 
\ion{He}{i}$^*$ originates from an \ion{H}{ii} region, the balmer absorption in
the associated partially ionized zone/photodissociation zone, and the extinction 
is caused by a very dense distribution of dust most likely associated with the 
gas causing the metal and molecular line absorption (e.g., \ion{Ca}{i}, \ion{Ca}{ii},
CH$^+$). The closest 
local analog to the region causing the dust extinction is probably the sightline to 
SN\,2014J in in the starburst galaxy M82: 
along this sightline similar extinction is seen \citep{Amanullah14}. Furthermore,
similar absorption line properties are also seen, e.g. strong Calcium absorption and
absorption from CH$^+$ \citep{Ritchey14}. Sightlines like these may be responsible for
turning some bursts into dark bursts. 

It is somewhat uncomfortable that the two peculiarities of the sightline
towards GRB\,140506A do not seem to have a common explanation, but to the best
of our judgement this is how the situation seems to be. Whatever the correct
interpretation of the steep UV extinction is, it is promising that the
phenomenon apparently is seen in several other types of objects, i.e. AGN
\citep{Fynbo13,Jiang13,Karen14}, SNe \citep{Goobar08,Amanullah14} and towards
the centre of M31 \citep{Dong14}. The outlook for a definitive solution to the
mystery is hence good.

\begin{acknowledgements}
We are grateful for a very constructive report from an anonymous referee.
We are also grateful for helpful discussions with Ariel Goobar, Rene Oudmaijer,
Moire Prescott, Kristian Finlator, Sebastian H{\"o}nig and J. Xavier Procaska.
We are also grateful to the wider X-shooter GRB follow-up team and the staff at
the La Silla Paranal Observatory without which this kind of research would not
be possible.  The Dark Cosmology Centre is funded by the DNRF. The research
leading to these results has received funding from the European Research
Council under the European Union's Seventh Framework Program
(FP7/2007-2013)/ERC Grant agreement no.  EGGS-278202. S. Schulze acknowledges
support from CONICYT-Chile FONDECYT 3140534, Basal-CATA PFB-06/2007, and
Project IC120009 "Millennium Institute of Astrophysics (MAS)" of Iniciativa
Cient\'{\i}fica Milenio del Ministerio de Econom\'{\i}a, Fomento y Turismo.
The research activity of Ad.U.P. and C.T. is supported by Spanish research
project AYA2012-39362-C02-02. Ad.U.P. acknowledges support by the European
Commission under the Marie Curie Career Integration Grant programme
(FP7-PEOPLE-2012-CIG 32230). Part of the funding for GROND (both hardware as
well as personnel) was generously granted from the Leibniz-Prize to Prof. G.
Hasinger (DFG grant HA 1850/28-1). PS acknowledges support through the Sofja
Kovalevskaja Award from the Alexander von Humboldt Foundation of Germany. DM
acknowledges support from the Instrument center for Danish Astrophysics (IDA)

\end{acknowledgements}

\bibliographystyle{aa}

\begin{thebibliography}{82}
\expandafter\ifx\csname natexlab\endcsname\relax\def\natexlab#1{#1}\fi

\bibitem[{{Aihara} {et~al.}(2011){Aihara}, {Allende Prieto}, {An}, {Anderson},
  {Aubourg}, {Balbinot}, {Beers}, {Berlind}, {Bickerton}, {Bizyaev}, {Blanton},
  {Bochanski}, {Bolton}, {Bovy}, {Brandt}, {Brinkmann}, {Brown}, {Brownstein},
  {Busca}, {Campbell}, {Carr}, {Chen}, {Chiappini}, {Comparat}, {Connolly},
  {Cortes}, {Croft}, {Cuesta}, {da Costa}, {Davenport}, {Dawson}, {Dhital},
  {Ealet}, {Ebelke}, {Edmondson}, {Eisenstein}, {Escoffier}, {Esposito},
  {Evans}, {Fan}, {Femen{\'{\i}}a Castell{\'a}}, {Font-Ribera}, {Frinchaboy},
  {Ge}, {Gillespie}, {Gilmore}, {Gonz{\'a}lez Hern{\'a}ndez}, {Gott}, {Gould},
  {Grebel}, {Gunn}, {Hamilton}, {Harding}, {Harris}, {Hawley}, {Hearty}, {Ho},
  {Hogg}, {Holtzman}, {Honscheid}, {Inada}, {Ivans}, {Jiang}, {Johnson},
  {Jordan}, {Jordan}, {Kazin}, {Kirkby}, {Klaene}, {Knapp}, {Kneib},
  {Kochanek}, {Koesterke}, {Kollmeier}, {Kron}, {Lampeitl}, {Lang}, {Le Goff},
  {Lee}, {Lin}, {Long}, {Loomis}, {Lucatello}, {Lundgren}, {Lupton}, {Ma},
  {MacDonald}, {Mahadevan}, {Maia}, {Makler}, {Malanushenko}, {Malanushenko},
  {Mandelbaum}, {Maraston}, {Margala}, {Masters}, {McBride}, {McGehee},
  {McGreer}, {M{\'e}nard}, {Miralda-Escud{\'e}}, {Morrison}, {Mullally},
  {Muna}, {Munn}, {Murayama}, {Myers}, {Naugle}, {Neto}, {Nguyen}, {Nichol},
  {O'Connell}, {Ogando}, {Olmstead}, {Oravetz}, {Padmanabhan},
  {Palanque-Delabrouille}, {Pan}, {Pandey}, {P{\^a}ris}, {Percival},
  {Petitjean}, {Pfaffenberger}, {Pforr}, {Phleps}, {Pichon}, {Pieri}, {Prada},
  {Price-Whelan}, {Raddick}, {Ramos}, {Reyl{\'e}}, {Rich}, {Richards}, {Rix},
  {Robin}, {Rocha-Pinto}, {Rockosi}, {Roe}, {Rollinde}, {Ross}, {Ross},
  {Rossetto}, {S{\'a}nchez}, {Sayres}, {Schlegel}, {Schlesinger}, {Schmidt},
  {Schneider}, {Sheldon}, {Shu}, {Simmerer}, {Simmons}, {Sivarani}, {Snedden},
  {Sobeck}, {Steinmetz}, {Strauss}, {Szalay}, {Tanaka}, {Thakar}, {Thomas},
  {Tinker}, {Tofflemire}, {Tojeiro}, {Tremonti}, {Vandenberg}, {Vargas
  Maga{\~n}a}, {Verde}, {Vogt}, {Wake}, {Wang}, {Weaver}, {Weinberg}, {White},
  {White}, {Yanny}, {Yasuda}, {Yeche}, \& {Zehavi}}]{2011ApJS..193...29A}
{Aihara}, H., {Allende Prieto}, C., {An}, D., {et~al.} 2011, \apjs, 193, 29

\bibitem[{{Amanullah} {et~al.}(2014){Amanullah}, {Goobar}, {Johansson},
  {Banerjee}, {Venkataraman}, {Joshi}, {Ashok}, {Cao}, {Kasliwal}, {Kulkarni},
  {Nugent}, {Petrushevska}, \& {Stanishev}}]{Amanullah14}
{Amanullah}, R., {Goobar}, A., {Johansson}, J., {et~al.} 2014, \apjl, 788, L21

\bibitem[{{Bruzual} \& {Charlot}(2003)}]{bc03}
{Bruzual}, G. \& {Charlot}, S. 2003, \mnras, 344, 1000

\bibitem[{{Christensen} {et~al.}(2011){Christensen}, {Fynbo}, {Prochaska},
  {Th{\"o}ne}, {de Ugarte Postigo}, \& {Jakobsson}}]{Lise}
{Christensen}, L., {Fynbo}, J.~P.~U., {Prochaska}, J.~X., {et~al.} 2011, \apj,
  727, 73

\bibitem[{{Churchwell}(2002)}]{Churchwell02}
{Churchwell}, E. 2002, \araa, 40, 27

\bibitem[{{Covino} {et~al.}(2013){Covino}, {Melandri}, {Salvaterra}, {Campana},
  {Vergani}, {Bernardini}, {D'Avanzo}, {D'Elia}, {Fugazza}, {Ghirlanda},
  {Ghisellini}, {Gomboc}, {Jin}, {Kr{\"u}hler}, {Malesani}, {Nava},
  {Sbarufatti}, \& {Tagliaferri}}]{Covino13}
{Covino}, S., {Melandri}, A., {Salvaterra}, R., {et~al.} 2013, \mnras, 432,
  1231

\bibitem[{{De Cia} {et~al.}(2012){De Cia}, {Ledoux}, {Fox}, {Vreeswijk},
  {Smette}, {Petitjean}, {Bj{\"o}rnsson}, {Fynbo}, {Hjorth}, \&
  {Jakobsson}}]{DeCia12}
{De Cia}, A., {Ledoux}, C., {Fox}, A.~J., {et~al.} 2012, \aap, 545, A64

\bibitem[{{De Cia} {et~al.}(2013){De Cia}, {Ledoux}, {Savaglio}, {Schady}, \&
  {Vreeswijk}}]{DeCia}
{De Cia}, A., {Ledoux}, C., {Savaglio}, S., {Schady}, P., \& {Vreeswijk}, P.~M.
  2013, \aap, 560, A88

\bibitem[{{de Ugarte Postigo} {et~al.}(2012){de Ugarte Postigo}, {Fynbo},
  {Th{\"o}ne}, {Christensen}, {Gorosabel}, {Milvang-Jensen}, {Schulze},
  {Jakobsson}, {Wiersema}, {S{\'a}nchez-Ram{\'{\i}}rez}, {Leloudas}, {Zafar},
  {Malesani}, \& {Hjorth}}]{Antonio12}
{de Ugarte Postigo}, A., {Fynbo}, J.~P.~U., {Th{\"o}ne}, C.~C., {et~al.} 2012,
  \aap, 548, A11

\bibitem[{{D'Elia} {et~al.}(2007){D'Elia}, {Fiore}, {Meurs}, {Chincarini},
  {Melandri}, {Norci}, {Pellizza}, {Perna}, {Piranomonte}, {Sbordone},
  {Stella}, {Tagliaferri}, {Vergani}, {Ward}, {Angelini}, {Antonelli},
  {Burrows}, {Campana}, {Capalbi}, {Cimatti}, {Costa}, {Cusumano}, {Della
  Valle}, {Filliatre}, {Fontana}, {Frontera}, {Fugazza}, {Gehrels}, {Giannini},
  {Giommi}, {Goldoni}, {Guetta}, {Israel}, {Lazzati}, {Malesani}, {Marconi},
  {Mason}, {Mereghetti}, {Mirabel}, {Molinari}, {Moretti}, {Nousek}, {Perri},
  {Piro}, {Stratta}, {Testa}, \& {Vietri}}]{DElia07}
{D'Elia}, V., {Fiore}, F., {Meurs}, E.~J.~A., {et~al.} 2007, \aap, 467, 629

\bibitem[{{D'Elia} {et~al.}(2014){D'Elia}, {Fynbo}, {Goldoni}, {Covino}, {de
  Ugarte Postigo}, {Ledoux}, {Calura}, {Gorosabel}, {Malesani}, {Matteucci},
  {S{\'a}nchez-Ram{\'{\i}}rez}, {Savaglio}, {Castro-Tirado}, {Hartoog},
  {Kaper}, {Mu{\~n}oz-Darias}, {Pian}, {Piranomonte}, {Tagliaferri}, {Tanvir},
  {Vergani}, {Watson}, \& {Xu}}]{DElia14}
{D'Elia}, V., {Fynbo}, J.~P.~U., {Goldoni}, P., {et~al.} 2014, \aap, 564, A38

\bibitem[{{Dong} {et~al.}(2014){Dong}, {Li}, {Wang}, {Lauer}, {Olsen}, {Saha},
  {Dalcanton}, {Gordon}, {Fouesneau}, {Bell}, \& {Bianchi}}]{Dong14}
{Dong}, H., {Li}, Z., {Wang}, Q.~D., {et~al.} 2014, \apj, 785, 136

\bibitem[{{Dressler} {et~al.}(2011){Dressler}, {Bigelow}, {Hare}, {Sutin},
  {Thompson}, {Burley}, {Epps}, {Oemler}, {Bagish}, {Birk}, {Clardy},
  {Gunnels}, {Kelson}, {Shectman}, \& {Osip}}]{Dressler2011a}
{Dressler}, A., {Bigelow}, B., {Hare}, T., {et~al.} 2011, \pasp, 123, 288

\bibitem[{{El{\'{\i}}asd{\'o}ttir} {et~al.}(2009){El{\'{\i}}asd{\'o}ttir},
  {Fynbo}, {Hjorth}, {Ledoux}, {Watson}, {Andersen}, {Malesani}, {Vreeswijk},
  {Prochaska}, {Sollerman}, \& {Jaunsen}}]{Ardis09}
{El{\'{\i}}asd{\'o}ttir}, {\'A}., {Fynbo}, J.~P.~U., {Hjorth}, J., {et~al.}
  2009, \apj, 697, 1725

\bibitem[{{Ellison} {et~al.}(2006){Ellison}, {Vreeswijk}, {Ledoux}, {Willis},
  {Jaunsen}, {Wijers}, {Smette}, {Fynbo}, {M{\o}ller}, {Hjorth}, \&
  {Kaufer}}]{Ellison06}
{Ellison}, S.~L., {Vreeswijk}, P., {Ledoux}, C., {et~al.} 2006, \mnras, 372,
  L38

\bibitem[{{Evans} {et~al.}(2005){Evans}, {Smartt}, {Lee}, {Lennon}, {Kaufer},
  {Dufton}, {Trundle}, {Herrero}, {Sim{\'o}n-D{\'{\i}}az}, {de Koter},
  {Hamann}, {Hendry}, {Hunter}, {Irwin}, {Korn}, {Kudritzki}, {Langer},
  {Mokiem}, {Najarro}, {Pauldrach}, {Przybilla}, {Puls}, {Ryans}, {Urbaneja},
  {Venn}, \& {Villamariz}}]{Evans2005}
{Evans}, C.~J., {Smartt}, S.~J., {Lee}, J.-K., {et~al.} 2005, \aap, 437, 467

\bibitem[{{Ferland} {et~al.}(2013){Ferland}, {Porter}, {van Hoof}, {Williams},
  {Abel}, {Lykins}, {Shaw}, {Henney}, \& {Stancil}}]{ferland13}
{Ferland}, G.~J., {Porter}, R.~L., {van Hoof}, P.~A.~M., {et~al.} 2013, \rmxaa,
  49, 137

\bibitem[{{Fitzpatrick} \& {Massa}(2007)}]{FM07}
{Fitzpatrick}, E.~L. \& {Massa}, D. 2007, \apj, 663, 320

\bibitem[{{Friis} {et~al.}(2014){Friis}, {De Cia}, {Fynbo}, {Ledoux},
  {Vreeswijk}, {Malesani}, {Gorosabel}, {Starling}, {Jakobsson}, {Varela},
  {Watson}, \& {Wiersema}}]{Mette14}
{Friis}, M., {De Cia}, A., {Fynbo}, J.~P.~U., {et~al.} 2014, submitted to MNRAS

\bibitem[{{Fruchter} {et~al.}(2001){Fruchter}, {Krolik}, \&
  {Rhoads}}]{Fruchter01}
{Fruchter}, A., {Krolik}, J.~H., \& {Rhoads}, J.~E. 2001, \apj, 563, 597

\bibitem[{{Fynbo} {et~al.}(2007){Fynbo}, {Vreeswijk}, {Jakobsson}, {Jaunsen},
  {Ledoux}, {Malesani}, {Th{\"o}ne}, {Ellison}, {Gorosabel}, {Hjorth},
  {Jensen}, {Kouveliotou}, {Levan}, {M{\o}ller}, {Rol}, {Smette}, {Sollerman},
  {Starling}, {Tanvir}, {Watson}, {Wiersema}, {Wijers}, \& {Xu}}]{Fynbo07}
{Fynbo}, J., {Vreeswijk}, P., {Jakobsson}, P., {et~al.} 2007, The Messenger,
  130, 43

\bibitem[{{Fynbo} {et~al.}(2009){Fynbo}, {Jakobsson}, {Prochaska}, {Malesani},
  {Ledoux}, {de Ugarte Postigo}, {Nardini}, {Vreeswijk}, {Wiersema}, {Hjorth},
  {Sollerman}, {Chen}, {Th{\"o}ne}, {Bj{\"o}rnsson}, {Bloom}, {Castro-Tirado},
  {Christensen}, {De Cia}, {Fruchter}, {Gorosabel}, {Graham}, {Jaunsen},
  {Jensen}, {Kann}, {Kouveliotou}, {Levan}, {Maund}, {Masetti},
  {Milvang-Jensen}, {Palazzi}, {Perley}, {Pian}, {Rol}, {Schady}, {Starling},
  {Tanvir}, {Watson}, {Xu}, {Augusteijn}, {Grundahl}, {Telting}, \&
  {Quirion}}]{Fynbo09}
{Fynbo}, J.~P.~U., {Jakobsson}, P., {Prochaska}, J.~X., {et~al.} 2009, \apjs,
  185, 526

\bibitem[{{Fynbo} {et~al.}(2013){Fynbo}, {Krogager}, {Venemans}, {Noterdaeme},
  {Vestergaard}, {M{\o}ller}, {Ledoux}, \& {Geier}}]{Fynbo13}
{Fynbo}, J.~P.~U., {Krogager}, J.-K., {Venemans}, B., {et~al.} 2013, \apjs,
  204, 6

\bibitem[{{Fynbo} {et~al.}(2006){Fynbo}, {Starling}, {Ledoux}, {Wiersema},
  {Th{\"o}ne}, {Sollerman}, {Jakobsson}, {Hjorth}, {Watson}, {Vreeswijk},
  {M{\o}ller}, {Rol}, {Gorosabel}, {N{\"a}r{\"a}nen}, {Wijers},
  {Bj{\"o}rnsson}, {Castro Cer{\'o}n}, {Curran}, {Hartmann}, {Holland},
  {Jensen}, {Levan}, {Limousin}, {Kouveliotou}, {Nelemans}, {Pedersen},
  {Priddey}, \& {Tanvir}}]{2006A&A...451L..47F}
{Fynbo}, J.~P.~U., {Starling}, R.~L.~C., {Ledoux}, C., {et~al.} 2006, \aap,
  451, L47

\bibitem[{{Fynbo} {et~al.}(2014){Fynbo}, {Tanvir}, {Jakobsson}, {Xu},
  {Malesani}, \& {Milvang-Jensen}}]{2014GCN..16217...1F}
{Fynbo}, J.~P.~U., {Tanvir}, N.~R., {Jakobsson}, P., {et~al.} 2014, GRB
  Coordinates Network, 16217, 1

\bibitem[{{Fynbo} {et~al.}(2001){Fynbo}, {Jensen}, {Gorosabel}, {Hjorth},
  {Pedersen}, {M{\o}ller}, {Abbott}, {Castro-Tirado}, {Delgado}, {Greiner},
  {Henden}, {Magazz{\`u}}, {Masetti}, {Merlino}, {Masegosa}, {{\O}stensen},
  {Palazzi}, {Pian}, {Schwarz}, {Cline}, {Guidorzi}, {Goldsten}, {Hurley},
  {Mazets}, {McClanahan}, {Montanari}, {Starr}, \& {Trombka}}]{Fynbo01}
{Fynbo}, J.~U., {Jensen}, B.~L., {Gorosabel}, J., {et~al.} 2001, \aap, 369, 373

\bibitem[{{Goldoni} {et~al.}(2006){Goldoni}, {Royer}, {Fran{\c c}ois}, M.,
  {Blanc}, {Vernet}, {Modigliani}, \& {Larsen}}]{Goldoni06}
{Goldoni}, P., {Royer}, F., {Fran{\c c}ois}, P., {et~al.} 2006, in Presented at
  the Society of Photo-Optical Instrumentation Engineers (SPIE) Conference,
  Vol. 6269, Society of Photo-Optical Instrumentation Engineers (SPIE) Confere
  Series

\bibitem[{{Golenetskii} {et~al.}(2014){Golenetskii}, {Aptekar}, {Frederiks},
  {Pal'Shin}, {Oleynik}, {Ulanov}, {Svinkin}, {Tsvetkova}, {Lyssenko}, \&
  {Cline}}]{2014GCN..16223...1G}
{Golenetskii}, S., {Aptekar}, R., {Frederiks}, D., {et~al.} 2014, GRB
  Coordinates Network, 16223, 1

\bibitem[{{Gompertz} {et~al.}(2014){Gompertz}, {Burrows}, {Cenko}, {Holland},
  {Kennea}, {Krimm}, {Kuin}, {Lien}, {Mangano}, {Page}, \&
  {Siegel}}]{2014GCN..16214...1G}
{Gompertz}, B.~P., {Burrows}, D.~N., {Cenko}, S.~B., {et~al.} 2014, GRB
  Coordinates Network, 16214, 1

\bibitem[{{Goobar}(2008)}]{Goobar08}
{Goobar}, A. 2008, \apjl, 686, L103

\bibitem[{{Greiner} {et~al.}(2008){Greiner}, {Bornemann}, {Clemens}, {Deuter},
  {Hasinger}, {Honsberg}, {Huber}, {Huber}, {Krauss}, {Kr{\"u}hler},
  {K{\"u}pc{\"u} Yolda{\c s}}, {Mayer-Hasselwander}, {Mican}, {Primak},
  {Schrey}, {Steiner}, {Szokoly}, {Th{\"o}ne}, {Yolda{\c s}}, {Klose}, {Laux},
  \& {Winkler}}]{2008PASP..120..405G}
{Greiner}, J., {Bornemann}, W., {Clemens}, C., {et~al.} 2008, \pasp, 120, 405

\bibitem[{{Greiner} {et~al.}(2011){Greiner}, {Kr{\"u}hler}, {Klose}, {Afonso},
  {Clemens}, {Filgas}, {Hartmann}, {K{\"u}pc{\"u} Yolda{\c s}}, {Nardini},
  {Olivares E.}, {Rau}, {Rossi}, {Schady}, \& {Updike}}]{Greiner11}
{Greiner}, J., {Kr{\"u}hler}, T., {Klose}, S., {et~al.} 2011, \aap, 526, A30

\bibitem[{{Hollenbach} \& {Tielens}(1997)}]{HT97}
{Hollenbach}, D.~J. \& {Tielens}, A.~G.~G.~M. 1997, \araa, 35, 179

\bibitem[{{Jakobsson} {et~al.}(2006){Jakobsson}, {Fynbo}, {Ledoux},
  {Vreeswijk}, {Kann}, {Hjorth}, {Priddey}, {Tanvir}, {Reichart}, {Gorosabel},
  {Klose}, {Watson}, {Sollerman}, {Fruchter}, {de Ugarte Postigo}, {Wiersema},
  {Bj{\"o}rnsson}, {Chapman}, {Th{\"o}ne}, {Pedersen}, \&
  {Jensen}}]{Jakobsson06}
{Jakobsson}, P., {Fynbo}, J.~P.~U., {Ledoux}, C., {et~al.} 2006, \aap, 460, L13

\bibitem[{{Jakobsson} {et~al.}(2004{\natexlab{a}}){Jakobsson}, {Hjorth},
  {Fynbo}, {Watson}, {Pedersen}, {Bj{\"o}rnsson}, \& {Gorosabel}}]{Palli04}
{Jakobsson}, P., {Hjorth}, J., {Fynbo}, J.~P.~U., {et~al.} 2004{\natexlab{a}},
  \apjl, 617, L21

\bibitem[{{Jakobsson} {et~al.}(2004{\natexlab{b}}){Jakobsson}, {Hjorth},
  {Fynbo}, {Weidinger}, {Gorosabel}, {Ledoux}, {Watson}, {Bj{\"o}rnsson},
  {Gudmundsson}, {Wijers}, {M{\"o}ller}, {Pedersen}, {Sollerman}, {Henden},
  {Jensen}, {Gilmore}, {Kilmartin}, {Levan}, {Castro Cer{\'o}n},
  {Castro-Tirado}, {Fruchter}, {Kouveliotou}, {Masetti}, \&
  {Tanvir}}]{Jakobsson04}
{Jakobsson}, P., {Hjorth}, J., {Fynbo}, J.~P.~U., {et~al.} 2004{\natexlab{b}},
  \aap, 427, 785

\bibitem[{{Jenke}(2014)}]{2014GCN..16220...1J}
{Jenke}, P. 2014, GRB Coordinates Network, 16220, 1

\bibitem[{{Jiang} {et~al.}(2013){Jiang}, {Zhou}, {Ji}, {Shu}, {Liu}, {Wang},
  {Dong}, {Bai}, {Wang}, \& {Wang}}]{Jiang13}
{Jiang}, P., {Zhou}, H., {Ji}, T., {et~al.} 2013, \aj, 145, 157

\bibitem[{{Kennicutt}(1998)}]{Kennicutt}
{Kennicutt}, Jr., R.~C. 1998, \araa, 36, 189

\bibitem[{{Krongold} \& {Prochaska}(2013)}]{KP13}
{Krongold}, Y. \& {Prochaska}, J.~X. 2013, \apj, 774, 115

\bibitem[{{Kr{\"u}hler} {et~al.}(2011){Kr{\"u}hler}, {Greiner}, {Schady},
  {Savaglio}, {Afonso}, {Clemens}, {Elliott}, {Filgas}, {Gruber}, {Kann},
  {Klose}, {K{\"u}pc{\"u}-Yolda{\c s}}, {McBreen}, {Olivares}, {Pierini},
  {Rau}, {Rossi}, {Nardini}, {Nicuesa Guelbenzu}, {Sudilovsky}, \&
  {Updike}}]{Tom11}
{Kr{\"u}hler}, T., {Greiner}, J., {Schady}, P., {et~al.} 2011, \aap, 534, A108

\bibitem[{{Kr{\"u}hler} {et~al.}(2008){Kr{\"u}hler}, {K{\"u}pc{\"u} Yolda{\c
  s}}, {Greiner}, {Clemens}, {McBreen}, {Primak}, {Savaglio}, {Yolda{\c s}},
  {Szokoly}, \& {Klose}}]{Tom08}
{Kr{\"u}hler}, T., {K{\"u}pc{\"u} Yolda{\c s}}, A., {Greiner}, J., {et~al.}
  2008, \apj, 685, 376

\bibitem[{{Kr{\"u}hler} {et~al.}(2013){Kr{\"u}hler}, {Ledoux}, {Fynbo},
  {Vreeswijk}, {Schmidl}, {Malesani}, {Christensen}, {De Cia}, {Hjorth},
  {Jakobsson}, {Kann}, {Kaper}, {Vergani}, {Afonso}, {Covino}, {de Ugarte
  Postigo}, {D'Elia}, {Filgas}, {Goldoni}, {Greiner}, {Hartoog},
  {Milvang-Jensen}, {Nardini}, {Piranomonte}, {Rossi},
  {S{\'a}nchez-Ram{\'{\i}}rez}, {Schady}, {Schulze}, {Sudilovsky}, {Tanvir},
  {Tagliaferri}, {Watson}, {Wiersema}, {Wijers}, \& {Xu}}]{Kruhler13}
{Kr{\"u}hler}, T., {Ledoux}, C., {Fynbo}, J.~P.~U., {et~al.} 2013, \aap, 557,
  A18

\bibitem[{{Lazzati} \& {Perna}(2002)}]{LP02}
{Lazzati}, D. \& {Perna}, R. 2002, \mnras, 330, 383

\bibitem[{{Leighly} {et~al.}(2011){Leighly}, {Dietrich}, \&
  {Barber}}]{leighly11}
{Leighly}, K.~M., {Dietrich}, M., \& {Barber}, S. 2011, \apj, 728, 94

\bibitem[{{Leighly} {et~al.}(2014){Leighly}, {Terndrup}, {Baron}, {Lucy},
  {Dietrich}, \& {Gallagher}}]{Karen14}
{Leighly}, K.~M., {Terndrup}, D.~M., {Baron}, E., {et~al.} 2014, \apj, 788, 123

\bibitem[{{Li} {et~al.}(2008){Li}, {Liang}, {Kann}, {Wei}, {Klose}, \&
  {Wang}}]{Li08}
{Li}, A., {Liang}, S.~L., {Kann}, D.~A., {et~al.} 2008, \apj, 685, 1046

\bibitem[{{Markwardt} {et~al.}(2014){Markwardt}, {Barthelmy}, {Baumgartner},
  {Cummings}, {Gehrels}, {Gompertz}, {Krimm}, {Lien}, {Palmer}, {Sakamoto},
  {Stamatikos}, {Tueller}, \& {Ukwatta}}]{2014GCN..16218...1M}
{Markwardt}, C.~B., {Barthelmy}, S.~D., {Baumgartner}, W.~H., {et~al.} 2014,
  GRB Coordinates Network, 16218, 1

\bibitem[{{Modigliani} {et~al.}(2010){Modigliani}, {Goldoni}, {Royer},
  {Haigron}, {Guglielmi}, {Fran{\c c}ois}, {Horrobin}, {Bristow}, {Vernet},
  {Moehler}, {Kerber}, {Ballester}, {Mason}, \& {Christensen}}]{Modigliani10}
{Modigliani}, A., {Goldoni}, P., {Royer}, F., {et~al.} 2010, in Society of
  Photo-Optical Instrumentation Engineers (SPIE) Conference Series, Vol. 7737,
  Society of Photo-Optical Instrumentation Engineers (SPIE) Conference Series

\bibitem[{{Nataf} {et~al.}(2013){Nataf}, {Gould}, {Fouqu{\'e}}, {Gonzalez},
  {Johnson}, {Skowron}, {Udalski}, {Szyma{\'n}ski}, {Kubiak},
  {Pietrzy{\'n}ski}, {Soszy{\'n}ski}, {Ulaczyk}, {Wyrzykowski}, \&
  {Poleski}}]{Nataf13}
{Nataf}, D.~M., {Gould}, A., {Fouqu{\'e}}, P., {et~al.} 2013, \apj, 769, 88

\bibitem[{{Oudmaijer} {et~al.}(1997){Oudmaijer}, {Drew}, {Barlow}, {Crawford},
  \& {Proga}}]{Oudmaijer1997}
{Oudmaijer}, R.~D., {Drew}, J.~E., {Barlow}, M.~J., {Crawford}, I.~A., \&
  {Proga}, D. 1997, \mnras, 291, 110

\bibitem[{{Pei}(1992)}]{Pei1992}
{Pei}, Y.~C. 1992, \apj, 395, 130

\bibitem[{{Perley} {et~al.}(2010){Perley}, {Bloom}, {Klein}, {Covino},
  {Minezaki}, {Wo{\'z}niak}, {Vestrand}, {Williams}, {Milne}, {Butler},
  {Updike}, {Kr{\"u}hler}, {Afonso}, {Antonelli}, {Cowie}, {Ferrero},
  {Greiner}, {Hartmann}, {Kakazu}, {K{\"u}pc{\"u} Yolda{\c s}}, {Morgan},
  {Price}, {Prochaska}, \& {Yoshii}}]{Perley10}
{Perley}, D.~A., {Bloom}, J.~S., {Klein}, C.~R., {et~al.} 2010, \mnras, 406,
  2473

\bibitem[{{Perley} {et~al.}(2013){Perley}, {Levan}, {Tanvir}, {Cenko}, {Bloom},
  {Hjorth}, {Kr{\"u}hler}, {Filippenko}, {Fruchter}, {Fynbo}, {Jakobsson},
  {Kalirai}, {Milvang-Jensen}, {Morgan}, {Prochaska}, \&
  {Silverman}}]{Perley13}
{Perley}, D.~A., {Levan}, A.~J., {Tanvir}, N.~R., {et~al.} 2013, \apj, 778, 128

\bibitem[{{Prochaska} {et~al.}(2006){Prochaska}, {Chen}, \&
  {Bloom}}]{Prochaska06}
{Prochaska}, J.~X., {Chen}, H.-W., \& {Bloom}, J.~S. 2006, \apj, 648, 95

\bibitem[{{Prochaska} {et~al.}(2007){Prochaska}, {Chen}, {Dessauges-Zavadsky},
  \& {Bloom}}]{2007ApJ...666..267P}
{Prochaska}, J.~X., {Chen}, H.-W., {Dessauges-Zavadsky}, M., \& {Bloom}, J.~S.
  2007, \apj, 666, 267

\bibitem[{{Prochaska} {et~al.}(2008){Prochaska}, {Dessauges-Zavadsky},
  {Ramirez-Ruiz}, \& {Chen}}]{Prochaska08}
{Prochaska}, J.~X., {Dessauges-Zavadsky}, M., {Ramirez-Ruiz}, E., \& {Chen},
  H.-W. 2008, \apj, 685, 344

\bibitem[{{Prochaska} {et~al.}(2009){Prochaska}, {Sheffer}, {Perley}, {Bloom},
  {Lopez}, {Dessauges-Zavadsky}, {Chen}, {Filippenko}, {Ganeshalingam}, {Li},
  {Miller}, \& {Starr}}]{2009ApJ...691L..27P}
{Prochaska}, J.~X., {Sheffer}, Y., {Perley}, D.~A., {et~al.} 2009, \apjl, 691,
  L27

\bibitem[{{Ritchey} {et~al.}(2014){Ritchey}, {Welty}, {Dahlstrom}, \&
  {York}}]{Ritchey14}
{Ritchey}, A.~M., {Welty}, D.~E., {Dahlstrom}, J.~A., \& {York}, D.~G. 2014,
  ArXiv e-prints

\bibitem[{{Rossi} {et~al.}(2012){Rossi}, {Klose}, {Ferrero}, {Greiner},
  {Arnold}, {Gonsalves}, {Hartmann}, {Updike}, {Kann}, {Kr{\"u}hler},
  {Palazzi}, {Savaglio}, {Schulze}, {Afonso}, {Amati}, {Castro-Tirado},
  {Clemens}, {Filgas}, {Gorosabel}, {Hunt}, {K{\"u}pc{\"u} Yolda{\c s}},
  {Masetti}, {Nardini}, {Nicuesa Guelbenzu}, {Olivares}, {Pian}, {Rau},
  {Schady}, {Schmidl}, {Yolda{\c s}}, \& {de Ugarte Postigo}}]{Rossi12}
{Rossi}, A., {Klose}, S., {Ferrero}, P., {et~al.} 2012, \aap, 545, A77

\bibitem[{{Savaglio}(2006)}]{2006NJPh....8..195S}
{Savaglio}, S. 2006, New Journal of Physics, 8, 195

\bibitem[{{Schady} {et~al.}(2012){Schady}, {Dwelly}, {Page}, {Kr{\"u}hler},
  {Greiner}, {Oates}, {de Pasquale}, {Nardini}, {Roming}, {Rossi}, \&
  {Still}}]{Schady12}
{Schady}, P., {Dwelly}, T., {Page}, M.~J., {et~al.} 2012, \aap, 537, A15

\bibitem[{{Sciortino} {et~al.}(1990){Sciortino}, {Vaiana}, {Harnden},
  {Ramella}, {Morossi}, {Rosner}, \& {Schmitt}}]{sciortino90}
{Sciortino}, S., {Vaiana}, G.~S., {Harnden}, Jr., F.~R., {et~al.} 1990, \apj,
  361, 621

\bibitem[{{Siegel} \& {Gompertz}(2014)}]{Siegel}
{Siegel}, M.~H. \& {Gompertz}, B.~P. 2014, GRB Coordinates Network, 16219, 1

\bibitem[{{Skrutskie} {et~al.}(2006){Skrutskie}, {Cutri}, {Stiening},
  {Weinberg}, {Schneider}, {Carpenter}, {Beichman}, {Capps}, {Chester},
  {Elias}, {Huchra}, {Liebert}, {Lonsdale}, {Monet}, {Price}, {Seitzer},
  {Jarrett}, {Kirkpatrick}, {Gizis}, {Howard}, {Evans}, {Fowler}, {Fullmer},
  {Hurt}, {Light}, {Kopan}, {Marsh}, {McCallon}, {Tam}, {Van Dyk}, \&
  {Wheelock}}]{2006AJ....131.1163S}
{Skrutskie}, M.~F., {Cutri}, R.~M., {Stiening}, R., {et~al.} 2006, \aj, 131,
  1163

\bibitem[{{Smoker} {et~al.}(2014){Smoker}, {Ledoux}, {Jehin}, {Keenan},
  {Kennedy}, {Cabanac}, \& {Melo}}]{Smoker14}
{Smoker}, J., {Ledoux}, C., {Jehin}, E., {et~al.} 2014, \mnras, 438, 1127

\bibitem[{{Sparre} {et~al.}(2014){Sparre}, {Hartoog}, {Kr{\"u}hler}, {Fynbo},
  {Watson}, {Wiersema}, {D'Elia}, {Zafar}, {Afonso}, {Covino}, {de Ugarte
  Postigo}, {Flores}, {Goldoni}, {Greiner}, {Hjorth}, {Jakobsson}, {Kaper},
  {Klose}, {Levan}, {Malesani}, {Milvang-Jensen}, {Nardini}, {Piranomonte},
  {Sollerman}, {S{\'a}nchez-Ram{\'{\i}}rez}, {Schulze}, {Tanvir}, {Vergani}, \&
  {Wijers}}]{Sparre14}
{Sparre}, M., {Hartoog}, O.~E., {Kr{\"u}hler}, T., {et~al.} 2014, \apj, 785,
  150

\bibitem[{{Starling} {et~al.}(2005){Starling}, {Wijers}, {Hughes}, {Tanvir},
  {Vreeswijk}, {Rol}, \& {Salamanca}}]{Starling05}
{Starling}, R.~L.~C., {Wijers}, R.~A.~M.~J., {Hughes}, M.~A., {et~al.} 2005,
  \mnras, 360, 305

\bibitem[{{Th{\"o}ne} {et~al.}(2013){Th{\"o}ne}, {Fynbo}, {Goldoni}, {de
  Ugarte}, {Campana}, {Vergani}, {Covino}, {Kr{\"u}hler}, {Kaper}, {Tanvir},
  {Zafar}, {D'Elia}, {Gorosabel}, {Greiner}, {Groot}, {Hammer}, {Jakobsson},
  {Klose}, {Levan}, {Milvang-Jensen}, {Nicuesa}, {Palazzi}, {Piranomonte},
  {Tagliaferri}, {Watson}, {Wiersema}, \& {Wijers}}]{cct13}
{Th{\"o}ne}, C.~C., {Fynbo}, J.~P.~U., {Goldoni}, P., {et~al.} 2013, \mnras,
  428, 3590

\bibitem[{{Tody}(1993)}]{Tody1993}
{Tody}, D. 1993, in Astronomical Society of the Pacific Conference Series,
  Vol.~52, Astronomical Data Analysis Software and Systems II, ed. R.~J.
  {Hanisch}, R.~J.~V. {Brissenden}, \& J.~{Barnes}, 173

\bibitem[{{Vernet} {et~al.}(2011){Vernet}, {Dekker}, {D'Odorico}, {Kaper},
  {Kjaergaard}, {Hammer}, {Randich}, {Zerbi}, {Groot}, {Hjorth}, {Guinouard},
  {Navarro}, {Adolfse}, {Albers}, {Amans}, {Andersen}, {Andersen}, {Binetruy},
  {Bristow}, {Castillo}, {Chemla}, {Christensen}, {Conconi}, {Conzelmann},
  {Dam}, {de Caprio}, {de Ugarte Postigo}, {Delabre}, {di Marcantonio},
  {Downing}, {Elswijk}, {Finger}, {Fischer}, {Flores}, {Fran{\c c}ois},
  {Goldoni}, {Guglielmi}, {Haigron}, {Hanenburg}, {Hendriks}, {Horrobin},
  {Horville}, {Jessen}, {Kerber}, {Kern}, {Kiekebusch}, {Kleszcz}, {Klougart},
  {Kragt}, {Larsen}, {Lizon}, {Lucuix}, {Mainieri}, {Manuputy}, {Martayan},
  {Mason}, {Mazzoleni}, {Michaelsen}, {Modigliani}, {Moehler}, {M{\o}ller},
  {Norup S{\o}rensen}, {N{\o}rregaard}, {P{\'e}roux}, {Patat}, {Pena}, {Pragt},
  {Reinero}, {Rigal}, {Riva}, {Roelfsema}, {Royer}, {Sacco}, {Santin},
  {Schoenmaker}, {Spano}, {Sweers}, {Ter Horst}, {Tintori}, {Tromp}, {van
  Dael}, {van der Vliet}, {Venema}, {Vidali}, {Vinther}, {Vola}, {Winters},
  {Wistisen}, {Wulterkens}, \& {Zacchei}}]{Vernet11}
{Vernet}, J., {Dekker}, H., {D'Odorico}, S., {et~al.} 2011, \aap, 536, A105

\bibitem[{{Vreeswijk} {et~al.}(2013){Vreeswijk}, {Ledoux}, {Raassen}, {Smette},
  {De Cia}, {Wo{\'z}niak}, {Fox}, {Vestrand}, \& {Jakobsson}}]{Vreeswijk13}
{Vreeswijk}, P.~M., {Ledoux}, C., {Raassen}, A.~J.~J., {et~al.} 2013, \aap,
  549, A22

\bibitem[{{Vreeswijk} {et~al.}(2007){Vreeswijk}, {Ledoux}, {Smette}, {Ellison},
  {Jaunsen}, {Andersen}, {Fruchter}, {Fynbo}, {Hjorth}, {Kaufer}, {M{\o}ller},
  {Petitjean}, {Savaglio}, \& {Wijers}}]{Vreeswijk07}
{Vreeswijk}, P.~M., {Ledoux}, C., {Smette}, A., {et~al.} 2007, \aap, 468, 83

\bibitem[{{Vreeswijk} {et~al.}(2011){Vreeswijk}, {Ledoux}, {Smette}, {Ellison},
  {Jaunsen}, {Andersen}, {Fruchter}, {Fynbo}, {Hjorth}, {Kaufer}, {M{\o}ller},
  {Petitjean}, {Savaglio}, \& {Wijers}}]{Vreeswijk11}
{Vreeswijk}, P.~M., {Ledoux}, C., {Smette}, A., {et~al.} 2011, \aap, 532, C3

\bibitem[{{Wang}(2005)}]{Wang05}
{Wang}, L. 2005, \apjl, 635, L33

\bibitem[{{Watson}(2009)}]{Watson09}
{Watson}, D. 2009, in Astronomical Society of the Pacific Conference Series,
  Vol. 414, Cosmic Dust - Near and Far, ed. T.~{Henning}, E.~{Gr{\"u}n}, \&
  J.~{Steinacker}, 277

\bibitem[{{Watson} {et~al.}(2006){Watson}, {Fynbo}, {Ledoux}, {Vreeswijk},
  {Hjorth}, {Smette}, {Andersen}, {Aoki}, {Augusteijn}, {Beardmore}, {Bersier},
  {Castro Cer{\'o}n}, {D'Avanzo}, {Diaz-Fraile}, {Gorosabel}, {Hirst},
  {Jakobsson}, {Jensen}, {Kawai}, {Kosugi}, {Laursen}, {Levan}, {Masegosa},
  {N{\"a}r{\"a}nen}, {Page}, {Pedersen}, {Pozanenko}, {Reeves}, {Rumyantsev},
  {Shahbaz}, {Sharapov}, {Sollerman}, {Starling}, {Tanvir}, {Torstensson}, \&
  {Wiersema}}]{Watson06}
{Watson}, D., {Fynbo}, J.~P.~U., {Ledoux}, C., {et~al.} 2006, \apj, 652, 1011

\bibitem[{{Watson} {et~al.}(2013){Watson}, {Zafar}, {Andersen}, {Fynbo},
  {Gorosabel}, {Hjorth}, {Jakobsson}, {Kr{\"u}hler}, {Laursen}, {Leloudas}, \&
  {Malesani}}]{Darach13}
{Watson}, D., {Zafar}, T., {Andersen}, A.~C., {et~al.} 2013, \apj, 768, 23

\bibitem[{{Wijers} \& {Galama}(1999)}]{WG99}
{Wijers}, R.~A.~M.~J. \& {Galama}, T.~J. 1999, \apj, 523, 177

\bibitem[{{Xiang} {et~al.}(2011){Xiang}, {Li}, \& {Zhong}}]{Xiang11}
{Xiang}, F.~Y., {Li}, A., \& {Zhong}, J.~X. 2011, \apj, 733, 91

\bibitem[{{Zafar} {et~al.}(2011){Zafar}, {Watson}, {Fynbo}, {Malesani},
  {Jakobsson}, \& {de Ugarte Postigo}}]{2011A&A...532A.143Z}
{Zafar}, T., {Watson}, D., {Fynbo}, J.~P.~U., {et~al.} 2011, \aap, 532, A143

\bibitem[{{Zhang} \& {M{\'e}sz{\'a}ros}(2004)}]{Bing04}
{Zhang}, B. \& {M{\'e}sz{\'a}ros}, P. 2004, International Journal of Modern
  Physics A, 19, 2385

\end{thebibliography}

\end{document}